\documentclass[iop]{emulateapj}
\usepackage{graphicx,url}
\usepackage{color}

\def\xmm{{\it XMM-Newton}}
\def\suz{{\it Suzaku}}

\def\nustar{{\it NuSTAR}}
\def\xillver{{\tt xillver}}
\def\relxill{{\tt relxill}}
\def\relline{{\tt relline}}

\def\ec{$E_\mathrm{cut}$}

\shorttitle{Constraining the high-energy cutoff}
\shortauthors{Garc\'{\i}a \& et al.}

\begin{document}

\title{On estimating the high-energy cutoff in the x-ray spectra of black
  holes via \\ reflection spectroscopy}

\author{Javier~A.~Garc\'ia\altaffilmark{1}, Thomas~Dauser\altaffilmark{2},
  James~F.~Steiner\altaffilmark{1}, Jeffrey~E.~McClintock\altaffilmark{1},
  Mason~L.~Keck\altaffilmark{1,3}, J\"{o}rn Wilms\altaffilmark{2}}

\altaffiltext{1}{Harvard-Smithsonian Center for Astrophysics,
  60 Garden St., Cambridge, MA 02138 USA; javier@head.cfa.harvard.edu,
  jem@cfa.harvard.edu, jsteiner@head.cfa.harvard.edu}

\altaffiltext{2}{Dr.\ Karl Remeis-Observatory and Erlangen Centre for 
  Astroparticle Physics, Sternwartstr.~7, 96049 Bamberg, Germany;
  thomas.dauser@sternwarte.uni-erlangen.de}

\altaffiltext{3}{Institute for Astrophysical Research, Boston University, 725
  Commonwealth Avenue, Boston, MA 02215, USA;  \email{keckm@bu.edu}}

%

\begin{abstract}

  The fundamental parameters describing the coronal spectrum of an
  accreting black hole are the slope $\Gamma$ of the power-law continuum
  and the energy \ec\ at which it rolls over.  Remarkably, this latter
  parameter can be accurately measured for values as high as 1~MeV by
  modeling the spectrum of X-rays reflected from a black hole accretion
  disk at energies below $100$~keV. This is possible because the details
  in the reflection spectrum, rich in fluorescent lines and other atomic
  features, are very sensitive to the spectral shape of the hardest
  coronal radiation illuminating the disk.  We show that by fitting
  simultaneous \nustar\ (3--79~keV) and low-energy (e.g., \suz) data
  with the most recent version of our reflection model \relxill\ one can
  obtain reasonable constraints on \ec\ at energies from tens of keV up
  to 1~MeV, for a source as faint as 1~mCrab in a 100~ks observation.

\end{abstract}

\keywords{accretion, accretion disks -- atomic processes -- black hole physics}

%
%
%
%
\section{Introduction}\label{sec:intro}

X-ray reflection spectroscopy is currently our most effective tool for
probing accretion processes and space-time near the supermassive black
holes that power active galactic nuclei (AGN).  In the current paradigm,
the power-law continuum is produced either in a central hot corona
\citep[e.g., ][]{sak73,haa93} or at the base of a jet
\citep[e.g.,][]{mat92,mar05}. A fraction of this radiation illuminates
the accretion disk and is reprocessed into a rich reflection spectrum
consisting of fluorescent lines and other features, which provides
detailed information on the composition and ionization state of the
accretion disk and constraints on the structure of the corona.
Importantly, reflection spectroscopy of AGN is {\it the} way to infer
the spin of their black holes via modeling key spectral features, such
as the Fe K line ($\sim 6.4$--$6.9$~keV), the Fe K edge ($\sim
7$--$8$~keV), and the Compton hump ($\sim 20$--$40$~keV).  Observational
data suggest that in many cases the illuminating power-law continuum
cuts off abruptly in the hard X-ray band \citep[e.g.,][]{zdz00}. Most
X-ray detectors are unable to accurately constrain this cutoff energy
because they either have limited sensitivity at high-energies or none at
all. Thus, reflection models have been traditionally simplified by
adopting a fixed cutoff energy at a reasonable value, e.g.,
$E_{cut}=300$~keV \citep{ros05,gar10}.

The 2012 launch of \nustar\ \citep{har13} has revolutionized reflection
spectroscopy because its instruments provide both very low background 
and superb sensitivity over the band 3--79~keV, which captures all the 
key reflection features \citep[e.g.,][]{ris13,wal13,par14}.
The quality of the \nustar\ data demand reflection models
that properly treat the high-energy cutoff as a fit parameter.
We have developed advanced relativistic models of X-ray
reflection from ionized accretion disks; the most recent version is
\relxill\ \citep{gar14a}.  These models include a rich atomic
database, full treatment of the angular distribution of the reflected
radiation, and an improved geometrical model of the illuminating coronal
source.  Furthermore, the most recent version of \relxill\ also includes
the high-energy cutoff \ec\ as a fit parameter.  Likewise, \ec\ is
included as a free parameter in the most recent version of the
reflection code {\tt reflionx} \citep{ros05}.

In the standard picture, the power law spectrum is generated in the corona by
Compton up-scattering of thermal disk photons. The shape of the
spectrum depends on the geometry, temperature, and optical depth of the
corona \citep[e.g.,][]{ryb79}.  In particular, the
fractional energy change of a scattered photon (in the non-relativistic
limit)
\begin{equation}\label{eq:de}
\frac{\Delta E}{E} = \frac{4kT_e}{m_ec^2} - \frac{E}{m_ec^2}
\end{equation}
implies a cutoff of the power-law near $4kT_e$ because the energy
gained by the photon cannot exceed the energy of the electron
\citep[e.g.,][]{tit94}. In practice, \ec$\sim 2-3 kT_e$ due to
dispersion, geometrical, and relativistic effects \cite[e.g.,][]{pet01}.
Thus, an empirical estimate of \ec\ provides direct information on the
temperature $T_e$ of the coronal electrons.

The recent data obtained with \nustar\ has spotlighted \ec\
\citep[e.g.,][]{bre14,bal14,mar14,bal15}. In most of these studies, the
coverage of \nustar\ to 79~keV was sufficient to directly observe the
roll over of the power-law component ($E_{\rm cut} \sim 50-200$~keV).
However, fitting a \nustar\ spectrum of NGC~5506 with relativistic
reflection models, \cite{mat15} surprisingly found
\ec=$720^{+130}_{-190}$~keV. Likewise, in fitting a \suz+\nustar\ spectrum
of NGC~4151 \cite{kec15} found that a value of \ec\ near 1~MeV was
statistically required.

In this Letter, we demonstrate the remarkable ability of reflection
models to provide strong constraints on \ec\ up to several hundred keV,
values that greatly exceed the limit of the
detector bandpass (e.g., 79~keV for \nustar). While not detected
directly, this high-energy portion of the spectrum conditions the
ionization state and structure of the disk atmosphere, which in turn
modifies the observed reflection features.  Thus, by properly modeling
the reflection component at moderate energies, one can accurately
estimate \ec\ at extreme energies and thereby constrain the physical
properties of the corona.

%
%
\begin{figure*}
\centering
\includegraphics[scale=0.5,angle=0]{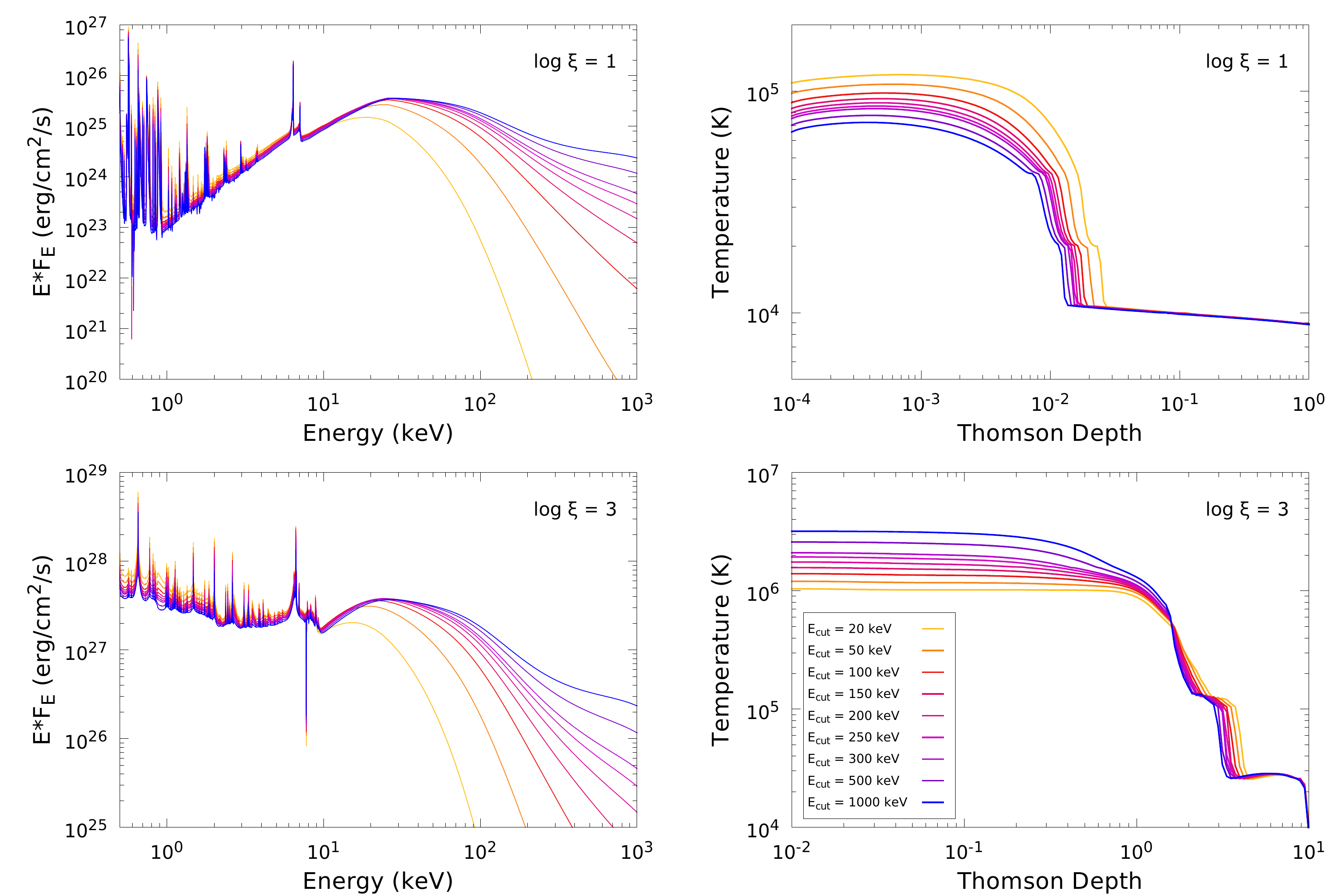}
\caption{The left panels show \xillver\ reflection spectra
  for log$\xi=1$ (top) and log$\xi=3$ (bottom) and for nine
  values of \ec\ (see legend in lower-right panel).  For all models, the
  illuminating spectrum is a power law with index $\Gamma=2$ and the
  iron abundance is solar. For simplicity we only show
  angle-averaged spectra \citep{gar14a}. The right panels show
  the temperature profile in the slab.}
\label{fig:xillver}
\end{figure*}

\section{Modeling the Reflection Spectrum}\label{sec:models}

The reflection model \relxill\ is a fusion of our
reflection code \xillver\ \citep{gar10,gar13a} and \relline\
\citep{dau10,dau13}, which is a general-relativistic ray tracing code.
We first describe the atomic physics part \xillver, which assumes
plane-parallel geometry and idealizes the disk as a slab with a total
optical depth of $\tau=10$ and a constant density of
$n=10^{15}$~cm$^{-3}$.  The illuminating source is assumed to have a
power-law spectrum with a photon index $\Gamma$ and an exponential
cutoff at high energies, \ec\footnote{For simplicity, we use a
	phenomenological e-folded power-law model rather than a proper model
  of thermal Comptonization such as that of \cite{zdz03}, a model whose
  accuracy has been demonstrated \citep[e.g.,][]{gie97,war02}.}.  The
intensity of the illumination is controlled by specifying the ionization
parameter:
\begin{equation}\label{eq:xi}
\xi = \frac{4\pi F_\mathrm{x}}{n},
\end{equation}
where $F_\mathrm{x}$ is the net ionizing flux in the energy band
1--1000~Ry.  A comprehensive grid of models has been calculated
and provided to the community for fitting observational data that cover
a wide range of model parameters. The most recent version of
\xillver\ includes models for several different values of the
high-energy cutoff, namely, $E_\mathrm{cut}=20, 50, 80, 100, 150, 200,
250, 300, 500, 1000$~keV.  In Figure~\ref{fig:xillver} we show examples
of models for these values of \ec\ for two ionization parameters,
log~$\xi=1$ and log~$\xi=3$.

As illustrated in Figure~\ref{fig:xillver} and discussed in
\cite{gar13a}, $E_\mathrm{cut}$ controls the intensity and shape
of the illuminating spectrum and thereby affects the global reflection
spectrum. The Compton hump is most dramatically affected, its peak
shifting to lower energies as \ec\ decreases. Weaker effects are seen at
lower energies ($\lesssim 10$~keV), which are mostly driven by changes
in the ionization state of the gas. As Equation~\ref{eq:xi} suggests,
two models with the same ionization
parameter (and hence the same $ F_\mathrm{x}$) but different values of
\ec\ will give rise to distinctive reflection spectra.  This
difference occurs because the illuminating spectrum with lower \ec\ will have a
significantly larger proportion of its photons concentrated at lower
energies, which favors photoelectric interactions over electron scattering
relative to the spectrum with high \ec.

Changes in \ec\ and the consequent shift in the balance of photons in
the hard and soft bands affects the ionization structure of the slab, as
illustrated in the two right panels of Figure~\ref{fig:xillver}.  On the
one hand, for low ionization the lower values of \ec\ (strong soft flux)
produce hotter atmospheres; as a consequence, in this regime the
temperature is mainly regulated by photoionization heating and
recombination cooling.  On the other hand, for high ionization the
hotter atmospheres correspond to higher values of \ec\ (strong hard
flux); in this regime, the temperature is controlled by electron
scattering of the abundant hard photons.  
Finally, we stress that
despite the opposing trends in the temperature profiles shown in
Figure~\ref{fig:xillver}, for a fixed value of $\xi$ the reflected
spectrum is always softer for smaller \ec.

\begin{figure}
\centering
\includegraphics[scale=0.4,angle=0]{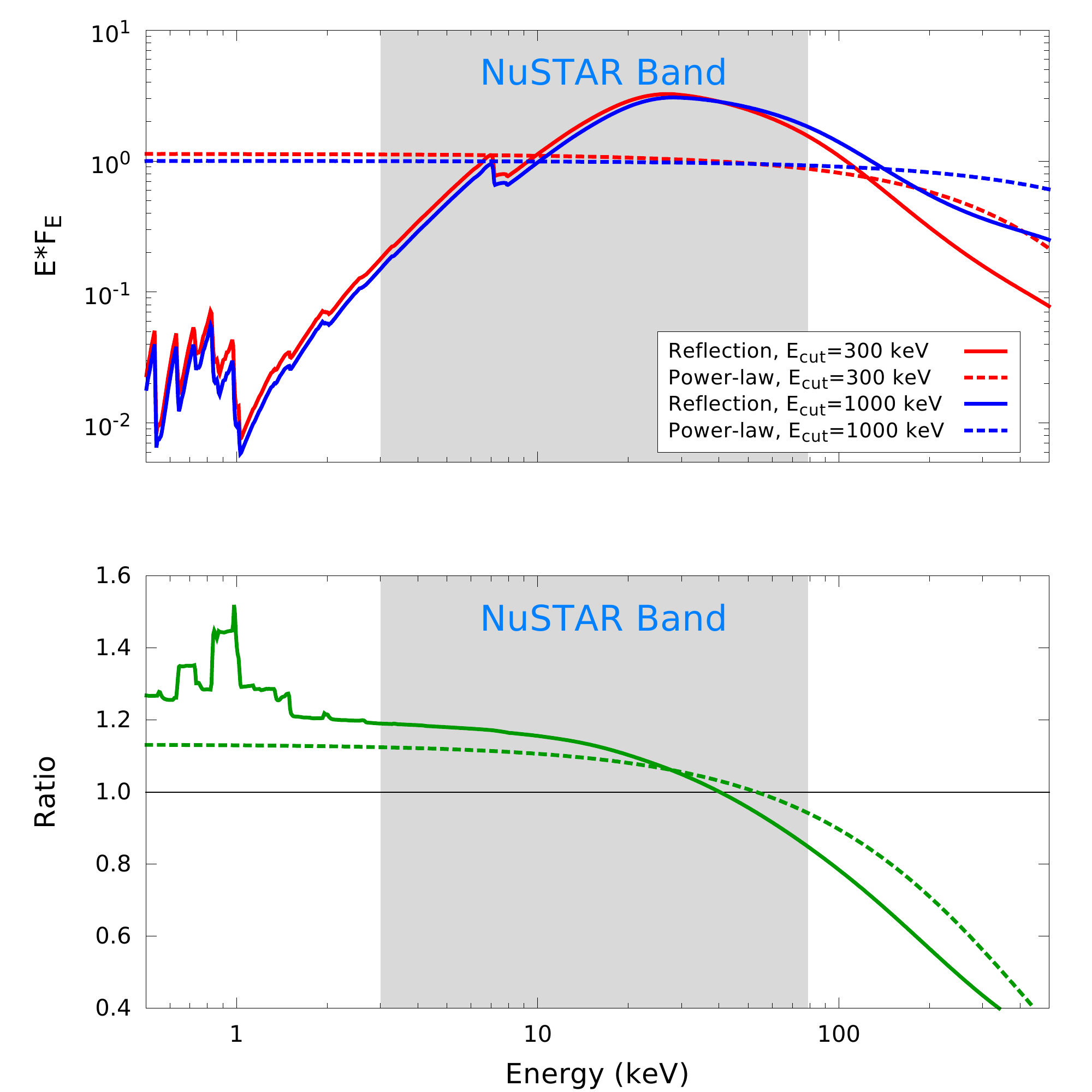}
\caption{ {\it Top:} Relativistically blurred reflection spectra (solid lines)
  computed using \relxill\ for log~$\xi=1$ and the respective power-law
  spectra (dashed lines) of the illuminating source with slope
  $\Gamma=2$. The red curves are for \ec=300~keV and the blue curves for
  \ec=1~MeV. {\it Bottom:} Ratio of the \ec=300~keV spectra to the
  \ec=1~MeV spectra. The ratio of the reflection/power-law spectra is
  plotted as a solid/dashed line.  The shaded box spans the \nustar\
  bandpass (3--79~keV). }
\label{fig:relxill}
\end{figure}

To produce the complete relativistic model \relxill, we convolve spectra
computed along the disk using \xillver\ with the general-relativistic
ray tracing code \relline\ \citep{dau10,dau13}.  The model \relxill\
self-consistently calculates the angle-dependent, relativistic ionized
reflection spectra \citep{gar14a} that are used in fitting data.
Figure~\ref{fig:relxill} shows reflected spectra (and the incident
power-law spectra) for log~$\xi=1$ calculated with \relxill.  For
clarity, we show only two models, for \ec=300~keV and \ec=1~MeV.  These
are the same as the corresponding two models shown in the top-left panel
of Figure~\ref{fig:xillver}, except that here the spectral features are
blurred by Doppler and gravitational-redshift effects.

The lower panel in Figure~\ref{fig:relxill} shows the ratio of the
\ec=300~keV model to the \ec=1~MeV model and also the ratio of the
incident power-law spectra.  In the \nustar\ band, while the two models
can be distinguished by considering either the power-law or the
reflected component, it is the stronger curvature of the reflected
component that is more telling.  At lower energies, in the \suz\ band,
the power-law is incapable of distinguishing between the two models
because the continuum is essentially identical (apart from the
normalization).  On the other hand, the reflected component shows
significant changes in spectral features, which sharply distinguish the
models.  Thus, whether one considers \nustar\ alone, or \nustar\ plus
\suz, it is possible to constrain \ec\ at energies far above the
bandpass of the detector, but in either case it is the reflected
component that carries the greater weight.

%
%

%
%
\begin{figure*}
\centering
\includegraphics[scale=0.5,angle=0]{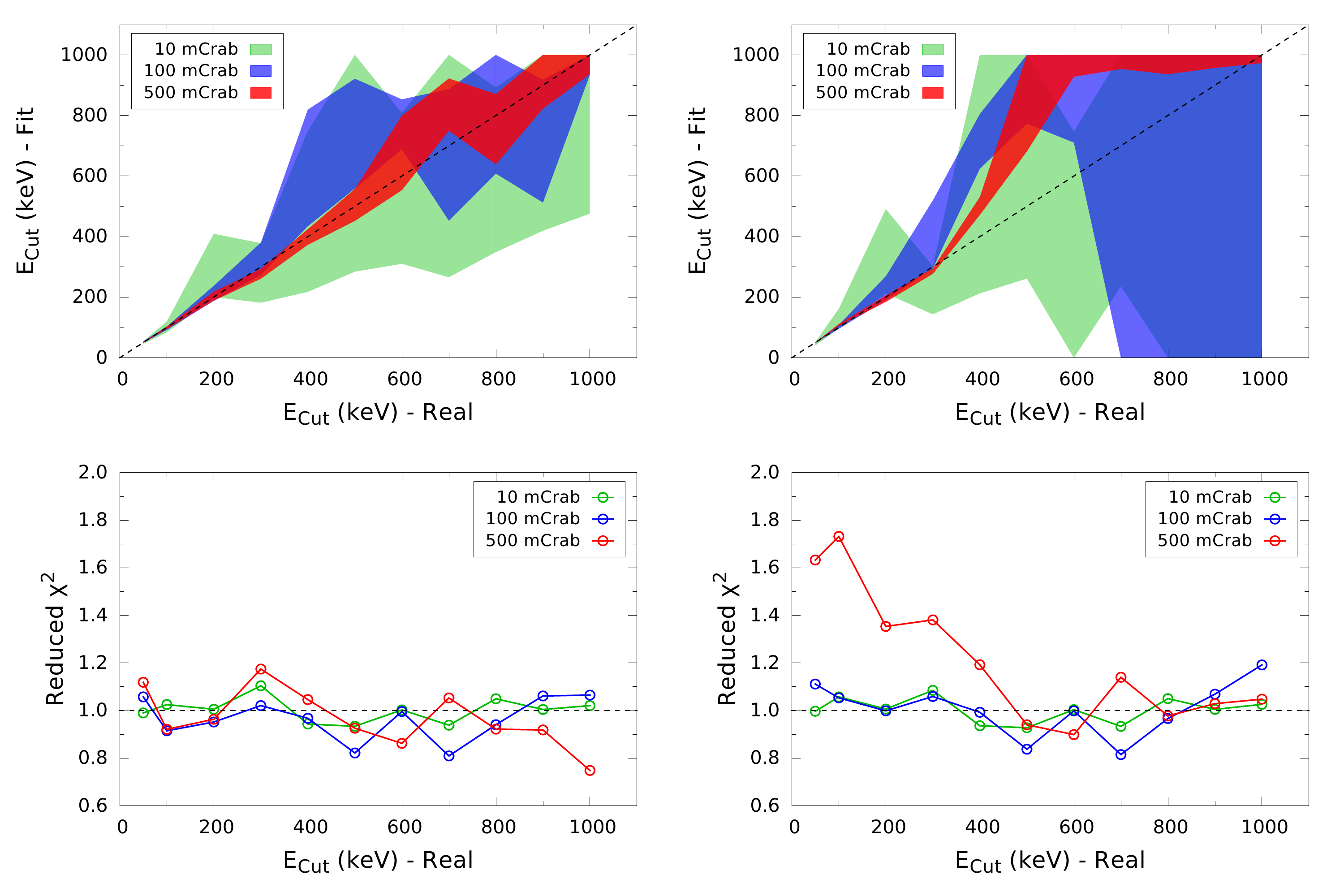}
\caption{Results of fitting simulated \nustar\ data. {\it Top:} Shown plotted are
the values of \ec\ returned by the fits compared to the values of \ec\
used in producing the simulated data for three assumed values of source
intensity (see legend).  Fits using Model~1 (left panel) demonstrate
that \ec\ can be accurately recovered for energies far above the 79~keV
limit of the \nustar\ bandpass.  Meantime, the very poor performance of
Model~2 (right panel) is obvious.  Error bars are at the 90\% level of
confidence.  {\it Bottom:} Goodness of fit; note the poor performance of
Model~2 at low to moderate values of \ec\ for the brightest source.
}
\label{fig:fits}
\end{figure*}

\section{Fits to Simulated \nustar\ Data}\label{sec:fits}

We first demonstrate that \nustar\ data alone can provide strong
constraints on \ec.  To this end, we simulate data using appropriate
response and background
files\footnote{\url{http://www.nustar.caltech.edu/page/response\_files}},
assuming a circular extraction region with 60$''$ radius centered 1$'$
off axis.  We combine the data from both \nustar\ telescopes (focal
plane modules FPMA and FPMB) and simulate spectra for an exposure time
of 100~ks.

We consider sources, both extragalactic and Galactic, with
intensities/fluxes that cover the range normally encountered in
practice.  Using the {\sc isis} package \citep{hou00}, we grouped the
data so that the minimum signal-to-noise ratio for each bin was 5 to
ensure the soundness of our $\chi^2$ analysis.  We additionally grouped
the data to ensure appropriate sampling, given that the resolution of
the \nustar\ detectors is 0.4/0.9~keV at 6/60~keV. Specifically, we
further binned the simulated data, which have uniform bin widths of
0.04~keV over the entire energy band, to 0.2~keV and 0.4~keV for
energies below and above 20~keV, respectively\footnote{We found that
  binning the data to a few samples per resolution element is important:
  Without this binning, the fitted values of \ec\ were consistently
  biased below the input.  This result suggests more generally that
  fitting oversampled data is not good practice.}.

In the simulations, we used the model {\tt Tbabs*relxill}
(Section~\ref{sec:models}), where {\tt Tbabs} models the interstellar
absorption \citep{wil00}.  We simulated spectra for eleven values of the
high-energy cutoff ranging from 50~keV to 1~MeV. The other model
parameters, which are the same for all simulations, are: the hydrogen
column density $N_\mathrm{H}=5\times10^{21}$~cm$^{-2}$; dimensionless
spin parameter $a=cJ/GM^2=0.998$; emissivity index $q=3$; inclination of
the disk $i=45$~deg; inner disk radius $R_\mathrm{in}=R_\mathrm{ISCO}$;
power-law $\Gamma=2$; ionization parameter $\log\xi=1$; iron abundance
$A_\mathrm{Fe}=1$; and the reflection fraction $R_f=3$, which is defined
as the ratio of the reflected to the power-law flux in the 20--40~keV
band.  Our calculations in this instance do not link the reflection
fraction to the emissivity index, which would require specifying the
geometry of the reflector and the illuminating source.

We emphasize that this Letter deals exclusively with
relativistically-blurred reflection by ionized gas in the very inner
portion of the accretion disk for which the reflection fraction can vary
widely, with values ranging from a few tenths to $>20$ for rapidly
spinning black holes \citep{dau14}.  Reflection in this regime is not to be
confused with reflection by distant, neutral material, which is observed
in the spectra of many sources and is usually modeled using the {\tt
  pexrav} model \citep[e.g.,][]{zdz99,zdz03}.
%

We generated the simulated spectra using the {\tt fakeit} task in {\sc
  xspec} v-12.8.2b \citep{arn96}, and we then fitted these spectra using
two different models, with \ec, $\Gamma$, $\xi$, $A_\mathrm{Fe}$, normalization, 
and $R_f$ free to vary:

{\bf Model~1: {\tt Tbabs*relxill}}. We first employ the same model used
to produce the simulated data in order to evaluate how well the
parameters, \ec\ in particular, can be recovered. 

{\bf Model~2: {\tt Tbabs*highecut*relxill}}. In this case, the parameter
\ec\ in \relxill\ is fixed at 1~MeV, while it is allowed to vary freely
in {\tt highecut}, a simple exponential convolution model acting on
\relxill.  Model~2 aims to test how accurately \ec\ is recovered when
the spectrum is described by modifying the high-energy region of an otherwise
incorrect reflection model (i.e., when ignoring the effects on the disk's
ionization structure described in \S~\ref{sec:models}).

Results of fitting the simulated data are shown in
Figure~\ref{fig:fits}.  The top panels show the fitted values of \ec\
versus the values used in simulating the data, with results for
Model~1/2 in the left/right panel.  The shaded regions indicate
uncertainties at the 90\% confidence level.  The colors are keyed to
source intensity.  The quality of the fits is indicated in the lower
panels.

Focusing on the results for Model~1 (top-left panel) and the 10~mCrab
source, which delivers $1.8\times10^6$~counts in the \nustar\ band, the
recovered values of \ec\ below 400 keV are of marginal quality, while
above that energy they only provide lower limits on \ec.  However, for
source intensities of $\sim 100-500$~mCrab ($\sim 10^7-10^8$~counts),
significantly better performance is achieved.  We conclude that for
high-quality data one can determine the high-energy cutoff with
reasonable precision using \nustar\ data alone.


Turning to the results for Model~2 (top-right panel), one immediately
sees the much poorer performance achieved in attempting to recover \ec\
by simply fitting the curvature in the power-law continuum component
using the {\tt highecut} model.  As mentioned in
Section~\ref{sec:intro}, this was the standard approach used before \ec\
was included as a fit parameter in reflection models.  The
{\tt highecut} model provides constraints of reasonable quality only for
\ec $<300$~keV. Furthermore, for the brightest source considered
(500~mCrab), Model~2 yields unsatisfactory results for \ec $>300$~keV
because not only is \ec\ significantly over-predicted, the error bars
are also misleadingly small. Interestingly, at lower energies
the quality of the fits are worse for the brightest source
(Figure~\ref{fig:fits}, bottom-right panel). Other model parameters are
also poorly constrained, e.g., the Fe abundance is grossly underestimated,
while both the power-law index and the reflection fraction $R_f$ are 
over-predicted.

Importantly, we note that Models 1 and 2 give the same quality of fit
(with the exception of the 500 mCrab simulation for low \ec) so that a
$\chi^2$-test does not generally promote one model over the other.  However, as we
have demonstrated, Model~2 returns incorrect estimates of the parameters
in all cases. In short, {\it good statistical precision does
 not ensure that the parameter values returned by a fit to a
  model are reliable or physically meaningful.}


\section{Including Low-Energy Data}\label{sec:suzaku}

%
%
\begin{figure*}
\centering
\includegraphics[width=\textwidth]{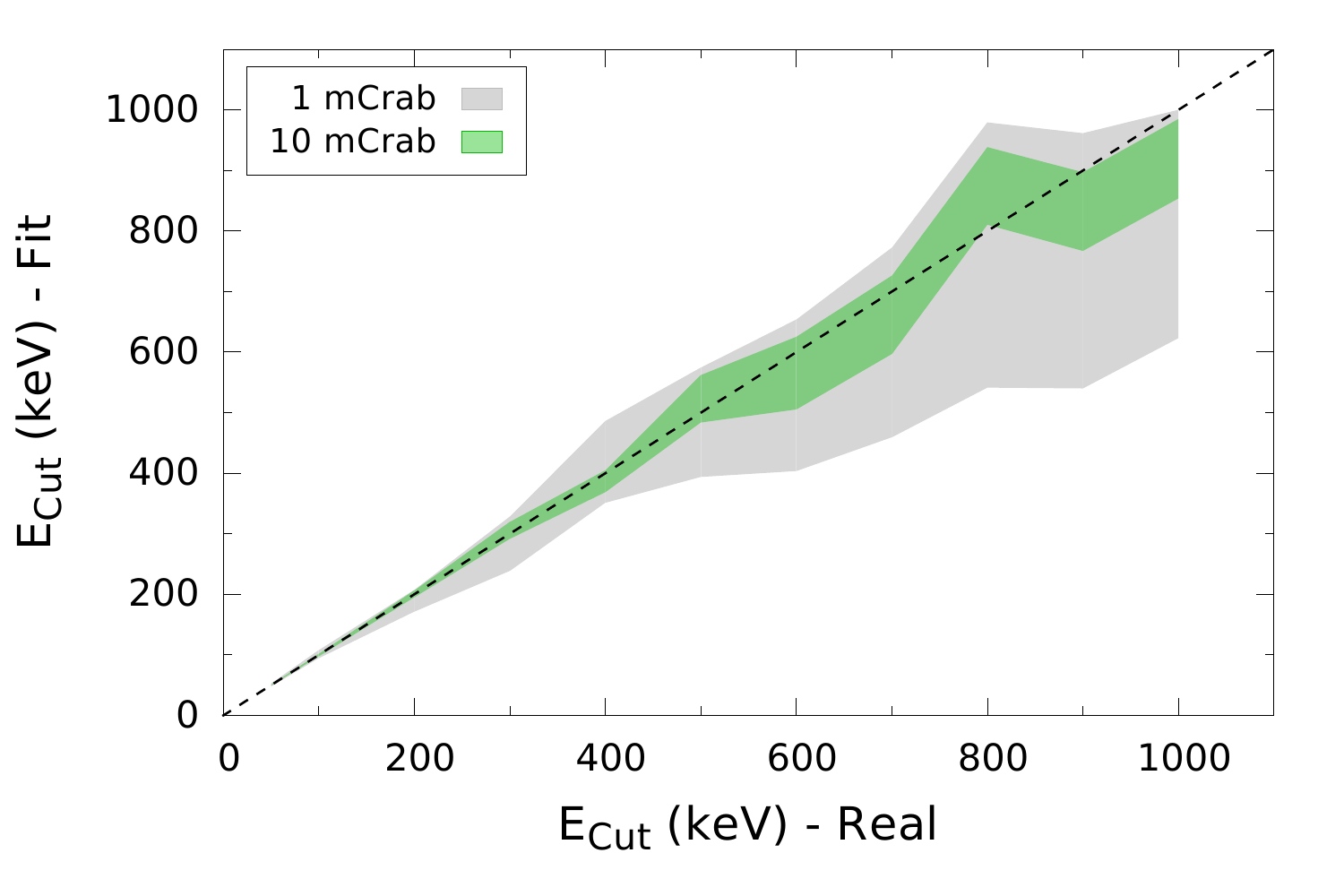}
\caption{Including low-energy \suz\ data, Model~1 delivers
  extraordinary performance in constraining \ec\ out to extreme energies
  compared to using \nustar\ data alone. The results shown
  here of fitting simulated data for a 10 mCrab source are superior to
  those obtained for a 500 mCrab source using only \nustar\ data (see
  the upper-left panel in Figure~\ref{fig:fits}).  }
\label{fig:suzaku}
\end{figure*}

Figure~\ref{fig:relxill} indicates that the low-energy component
($E\lesssim3$~keV) of the reflection spectrum is more sensitive to the
value of \ec\ than the component in the \nustar\ band.  Therefore, we
now consider spectra comprised of \suz\ plus \nustar\ data in the
expectation that this will significantly improve the constraints on \ec,
and the other parameters as well. The simulated low-energy data are
prepared using \suz\ response files, the same set of model parameters,
and the same source fluxes as those used in preparing the \nustar\ data.

In the following, we consider only Model~1. Fitting the combined set of
\suz\ (0.5--8~keV) and \nustar\ (3--79~keV) simulated spectra, we obtain
the outstanding result shown in Figure~\ref{fig:suzaku}. The fit for the
10 mCrab source is incomparably better than for the \nustar\ data alone
(compare upper-left panel of Figure~\ref{fig:fits}). The high-energy
cutoff is strongly constrained for all values of \ec\ considered; even
at 1~MeV, the error is only about 7\%. This excellent performance of our
reflection model is a consequence of the sensitivity of the reflection
component to changes in the high-energy portion of the illuminating
spectrum (Figure~\ref{fig:relxill}). These simulations clearly
demonstrate that fits to features in the relatively low-energy portion
of the spectrum, where the instruments are most sensitive, can deliver
accurate measurements of \ec\ at energies far beyond the limit of the
detector bandpass.  

Also shown in Figure~\ref{fig:suzaku} are the results for a 1~mCrab
source ($2\times10^5$~counts).  Remarkably, with the inclusion of
low-energy data even $\sim10^5$~counts are sufficient to recover \ec\
with reasonable precision; the error bars for \ec$>600$~keV are
$\sim30$\%.

\begin{figure*}
\centering
\includegraphics[width=\textwidth]{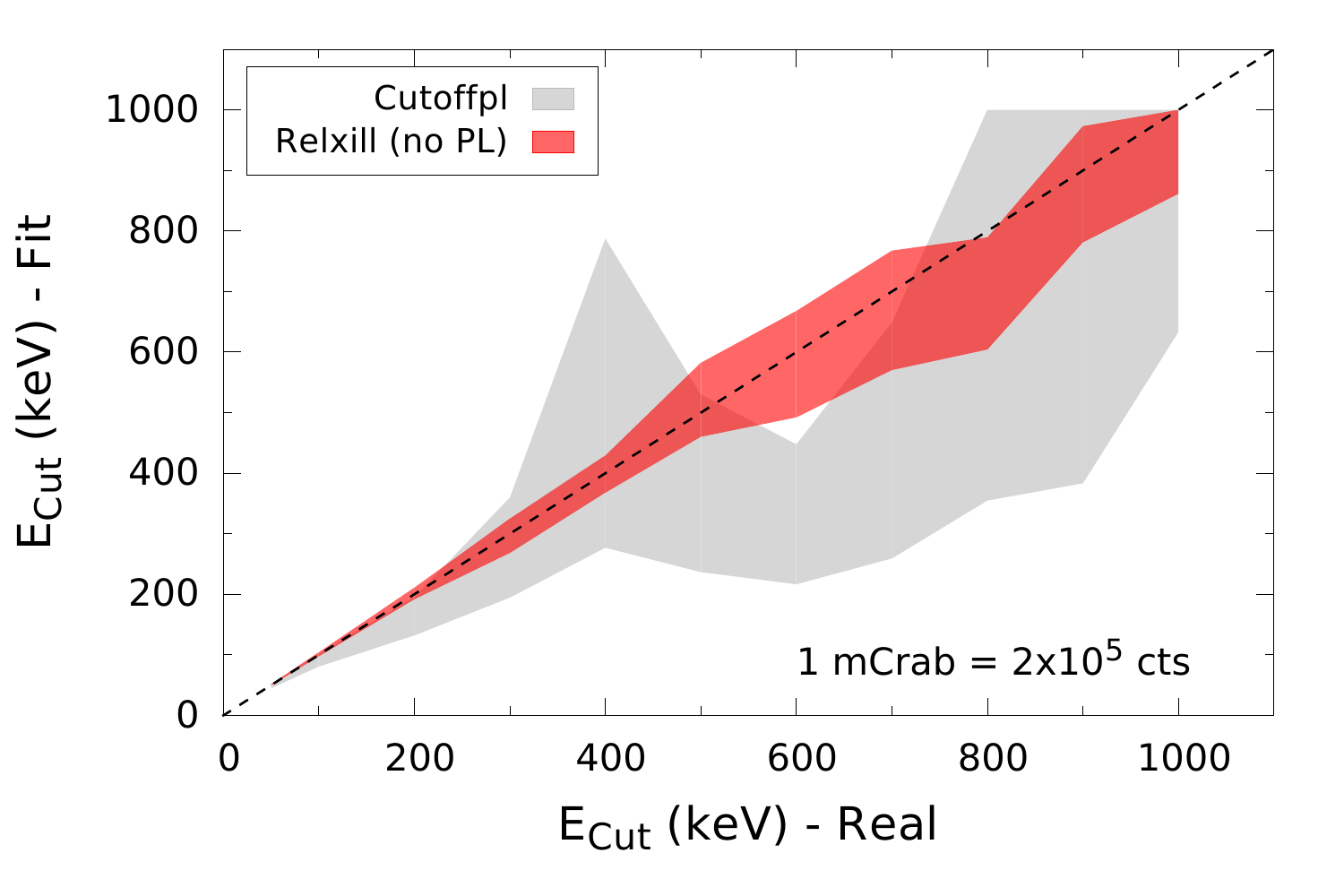}
\caption{ Results simulated for a 1 mCrab source demonstrating that a pure
reflection model (i.e., no power-law continuum) shown in red is far
superior to an e-folded power-law component (i.e., no reflection) in
constraining \ec. }
\label{fig:plcomp}
\end{figure*}

Finally, with two additional simulations we demonstrate that the
reflection component is much more important in constraining \ec\ than
the power law.  For the same 1 mCrab source, and including \suz\ data,
Figure~\ref{fig:plcomp} compares the performance of a cutoff power-law
model alone ({\tt cutoffpl}) with a pure reflection model.  These two
limiting cases can be compared directly to the intermediate case of our
adopted composite model (i.e., power law plus reflection) with $R_f=3$
shown in gray in Figure~\ref{fig:suzaku}.  The three cases considered
together show that stronger reflection translates into a better
constraint on \ec.

%
%
\section{Discussion and Conclusions}\label{sec:disco}

We have demonstrated that our reflection model \relxill\ can constrain
the physically important coronal parameter \ec\ for spectra with
intensities like those encountered in observations of AGN.  We have
furthermore shown that attempting to constrain \ec\ by imposing a cutoff
that is external to the reflection model yields relatively poor
constraints on \ec.  Such a simplified model inaccurately represents the
observed spectrum and introduces strong biases in the values inferred
for the key physical parameters.

We analyzed two simulated data sets using \relxill: (1) With \nustar\
data alone (3--79 keV), \ec$\gtrsim 400$~keV can only be constrained
if the spectrum contains $\gtrsim10^7$~counts,
which is infeasible for most observations of AGN.  (2) However, our
simulations indicate that for a simultaneous fit to \suz\ and \nustar\
data (0.5--79 keV), one can constrain \ec\ for a 1~mCrab source at
energies far above the limit of the detector bandpass.

Our results lead to two basic conclusions:

\begin{enumerate}

\item In constraining the cutoff energy, misleading results are obtained if
  the power-law component is cut off externally.  It is therefore
  essential to employ a detailed and accurate model of the reflection
  spectrum that includes \ec\ as an internal fit parameter.

\item Because the cutoff energy most strongly affects the low-energy
  part of the reflection spectrum, to obtain useful constraints on \ec\
  for AGN it is crucial to combine \nustar\ data with low-energy (e.g.,
  \suz\ or \xmm) data.

\end{enumerate}  

This limited study provides only a very cursory exploration of how
proper modeling of a reflection spectrum yields useful physical
constraints on the high-energy cutoff out to extreme energies.  A
thorough exploration of the subject is well beyond the scope of this
Letter.  For example, we illustrate the effects on the reflected
spectrum of varying \ec\ for the cases of log$\xi=1$ and 3
(Figure~\ref{fig:xillver}) while only presenting simulations for the former
case.  However, one expects that the effects will be quite different for
other choices of the ionization parameter. Moreover, the values of other
parameters, such as the reflection fraction, can affect the statistical
significance of the value returned for \ec.

%
%
%
\acknowledgments JG and JEM acknowledge the support of NASA grant
NNX11AD08G. JFS has been supported by NASA Hubble Fellowship 
HST-HF-51315.01. 
%
%
%
%
\bibliographystyle{apj}

%
%
%
%
\end{document}